# Electronic and atomic kinetics in solids irradiated with free-electron lasers or swift-heavy ions

N. Medvedev[*,1], A.E. Volkov[2,3,4], B. Ziaja[1,5]

[1]Center for Free-Electron Laser Science at DESY, Notkestr. 85, 22607 Hamburg, Germany;

[2]FLNR, JINR, Joliot-Curie 6, 141980 Dubna, Russia;

[3]NRC Kurchatov Institute, Kurchatov Sq. 1, 123182 Moscow, Russia;

[4]LPI of the Russian Academy of Sciences, Leninskij prospekt, 53,119991 Moscow, Russia

[5] Institute of Nuclear Physics, Polish Academy of Sciences, Radzikowskiego 152, Kraków 31-342, Poland

## Abstract

In this brief review we discuss the transient processes in solids under irradiation with femtosecond X-ray free-electron-laser (FEL) pulses and swift-heavy ions (SHI). Both kinds of irradiation produce highly excited electrons in a target on extremely short timescales. Transfer of the excess electronic energy into the lattice may lead to observable target modifications such as phase transitions and damage formation. Transient kinetics of material excitation and relaxation under FEL or SHI irradiation are comparatively discussed. The same origin for the electronic and atomic relaxation in both cases is demonstrated. Differences in these kinetics introduced by the geometrical effects (μm-size of a laser spot vs nm-size of an ion track) and initial irradiation (photoabsorption vs an ion impact) are analyzed. The basic mechanisms of electron transport and electron-lattice coupling are addressed. Appropriate models and their limitations are presented. Possibilities of thermal and nonthermal melting of materials under FEL and SHI irradiation are discussed.

**Keywords:** Free-electron laser; swift-heavy ion; electron kinetics; nonthermal melting; nonequilibrium

[*] Corresponding author: nikita.medvedev@desy.de

# I. Introduction

Systems far from equilibrium are actively investigated during the last decades (see, e.g., [1–7]). Effects detected in highly excited solids pose a number of challenges from both theoretical and experimental viewpoints [1]. For example, excitation by ultrafast energy deposition into the electronic subsystem of a material produces transient nonequilibrium states, which cannot be described by the standard techniques based on macroscopic quantities or local equilibrium conceptions [8–12].

Extreme initial excitation of the electronic subsystem of a solid is commonly produced in experiments by two ways.

Intense free-electron lasers, the 4-th generation light sources such as FLASH [13], LCLS [14], SACLA [15], FERMI [16], European XFEL [17], produce brilliant extreme ultraviolet and X-ray radiation with photon energies up to ~24 keV [17]. FELs pulse duration reaches a femtosecond timescale [14]. Their intensities are sufficient to trigger material modifications by a single shot. One of the most important advantages of femtosecond FELs is that the pulse duration is comparable with characteristic timescales of the basic processes in solids, e.g. the typical timescales of nonequilibrium electron cascades in material. A spot of X-ray FEL has a typical size of a μm. The photon attenuation length depends on the photon energy, and can be as short as a few tens of nanometers (for VUV at energy around the plasmon minimum), or as long as microns (for hard X-rays) [18,19]. Thus, material modifications produced by X-ray FEL are typically of a micron size.

Swift heavy ions (SHI, $M \geq 20m_p$, $m_p$ is the proton mass, $E > 1$ MeV/amu) decelerated in the electronic stopping regime excite, in the nanometric vicinities of their trajectories, primary electrons (so-called delta-electrons) up to energies similar to those of FEL-irradiation. For instance, UNILAC accelerator at GSI produces electrons with energies up to ~24 keV [20,21] by an impact of a nonrelativistic heavy ions with energies around the Bragg-peak realizing the electronic stopping [22]. SHI impacts can stimulate structure and phase transformations forming nanometric latent tracks along the SHI trajectories in a bulk. Their length can be of a few tens of microns or even longer, depending on the SHI energy and stopping power [22]. Surface damage in a shape of craters [23] and/or hillocks [24,25] can also be formed during an SHI irradiation.

In both cases, FEL and SHI irradiation, initially excited electrons scatter further elastically (exchanging kinetic energy with the atoms without secondary electron excitation) and inelastically (impact ionization of secondary electrons), ultimately relaxing to the low-energy states of the conduction band. Auger-decay of core holes brings them up to the valence band via a sequential cascade of events. Excess energy is later transferred to lattice atoms. These relaxation processes in the electron subsystem are physically the same for FEL or SHI irradiation. The differences are only introduced by two effects: (a) initial spectra of excited electrons that are different after photoabsorption and after an SHI impact; (b) geometry of the problem is different due to the difference in the spatial scales mentioned above. We will discuss these similarities and differences in the next sections.

The similarity of the physical processes during relaxation leads to mutual benefits for the communities investigating FEL and SHI irradiation. For example, pump-probe experiments studying transient electron kinetics are much easier to perform with FELs, since advanced synchronization schemes can achieve a femtosecond resolution [26,27]. The data on electron cascading in solids can then be utilized for modeling of SHI irradiation, which can be very useful considering that femtosecond resolution has not yet been achieved in SHI experiments.

A number of subsequent stages of nonequilibrium appear after ultrafast high-energy deposition into the electronic subsystem of a material:

(i) Initially excited electrons (on the femtosecond timescale) retain nonthermalized distribution until their thermalization (which may take up to a few hundred femtoseconds [8,11]).

During first ten femtoseconds [21,28,29], they are accompanied by the Auger cascades of core-holes, also being out of ionization equilibrium;

(ii) On the sub-picosecond timescale [9], the distribution of atomic energies in a solid under intense excitation may as well be nonthermal (non-Maxwellian);

(iii) Later, the electron ensemble may termalize at temperatures considerably higher than the lattice temperature. Electron-lattice coupling equilibrate these temperatures during a few to a few tens of picoseconds [1,2,8,30,31].

(iv) Additionally, effects of extensive spatial temperature and carrier density gradients can be important for the relaxation kinetics of the localized initial excitation, governing particle and energy transport [21,32–34].

Further material damage may proceed via a number of different channels: (a) thermal lattice heating by electron-phonon (or, more generally, electron-lattice) coupling; (b) nonthermal melting via modification of the interatomic potential caused by high excitation of the electron subsystem of a solid; (c) Coulomb explosion due to space-charge distribution resulting from transient spatial separation of excited electrons and lattice ions; (d) accumulation of structure defects during relaxation of electronic excitations, e.g. creation of color centers in alkali-halides due to creation and decay of self-trapping excitons. The contribution of these mechanisms depends on the peculiarities of the irradiation and target parameters.

In this paper we will only address the first two channels which realize in all targets, while the latter two are excluded from current considerations for the following reasons.

The channel "d" for electronic excitation is realized only for a number of wide band gap dielectrics (see e.g. [35], and [36] where the mechanism of point defect appearing due to creation and decay of self-trapped excitons is discussed). Also, multiple laser shots or prolonged radiation exposures are necessary for accumulation of point defect concentrations necessary for stimulation of lattice instability. Such multi-shot effects will not be discussed in this paper.

The Coulomb explosion (channel 'c') plays an important role only for finite systems such as molecules and nanoclusters: e.g. it was shown in Refs. [37,38] that increase of the size of a nanocluster results in decrease of the effect of Coulomb explosion for the interior of the cluster. The explosion then mostly affects its outer layer. Extrapolating, one can assume that in a bulk material Coulomb explosion does not occur, except for the near-surface region where the emitted electrons could create long-lasting charge non-neutrality. In this paper we will not consider such near-surface effects, focusing on the material deeply inside the bulk, far from a surface.

## II. Electron kinetics

At present, it is hardly possible to solve many-body time-dependent Schrödinger equation to trace the evolution of the highly-excited electron ensemble in a solid on the *ab-initio* level. Thus, for a calculation of nonequilibrium electron kinetics different approximations are used. In particular, one of the most popular is the one-particle semi-classical approximation resulting in Boltzmann kinetic equation for electrons [4,8,38–42]. Event-by-event Monte Carlo (MC) models [34,43,44] of independent electron scattering can be essentially considered as belonging to the same class of methods: introduction of correlations into the electron ensemble within the MC scheme (such as done e.g. in [11,45]) allows to include electron-electron interaction and Pauli's blocking effects similarly to the Boltzmann collision integral for electrons in solids.

The difference in the initial excitation of the electron subsystem by FEL or SHI appears, first of all, because photoabsorption is realized by quanta, whereas an ion impact excites the whole spectrum of electrons. However, the subsequent relaxation processes are similar. For example, in Fig. 1 the evolution of the total electron density is shown in both irradiation cases. Figure 1 shows evolution of the electron density in diamond irradiated with 1 fs FEL pulse, photon energy of 24 keV, fluence of 1 mJ/cm$^2$, vs irradiation with Ca ion with the energy of 457 MeV (11.4 MeV/amu). This particular ion energy is chosen for a better comparison of the two scenarios, because the maximum electron energy provided in the ion collision is ~24 keV i.e. the same as in the FEL case discussed.

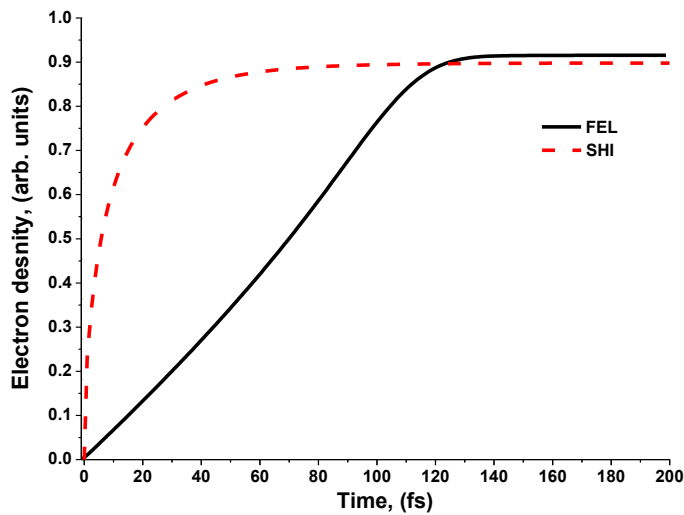

**Figure 1.** Evolution of the electron density in diamond irradiated with: an FEL pulse of 1 fs, photon energy 24 keV, fluence of 1 mJ/cm² (FEL), or Ca ion with energy 11.4 MeV/amu (SHI).

Differing timescales of the electron cascading can be observed before the saturation of the densities follows. The electron cascading started by photoabsorption at the energies around 24 keV excited with an FEL pulse takes around ~120 fs. After that, the excited electrons do not have sufficiently high energies to produce impact ionization, as will be discussed in the next section. In contrast, after the SHI impact, there is a lot of low-energy electrons in the first generation ionized by the ion which relax much faster (95% of electrons are produced already by 40 fs, 99% are excited by 90 fs). The relaxation of a minority of high-energy electrons still lasts a time comparable to that produced by an FEL pulse (~90 fs, as will become clear in the section II.2); these very few electrons, however, are highly energetic, thus, may be important to consider.

To emphasize the similarities and differences, below we will discuss the two cases in more detail.

### 1. FEL irradiation

Owing to a typical micron-size of a laser spot, homogeneous excitation can usually be assumed and periodic boundary conditions can be used for modeling. That significantly simplifies theoretical description, since zero-dimensional models could then be applied, tracing only evolution of the electron distribution in the energy space. XCASCADE[†] code [46], and its predecessors, were used to study evolution of electron distribution in different materials irradiated with an FEL pulse [11,45,47]. This code is also used to obtain the data presented in Fig. 1, and the results discussed below.

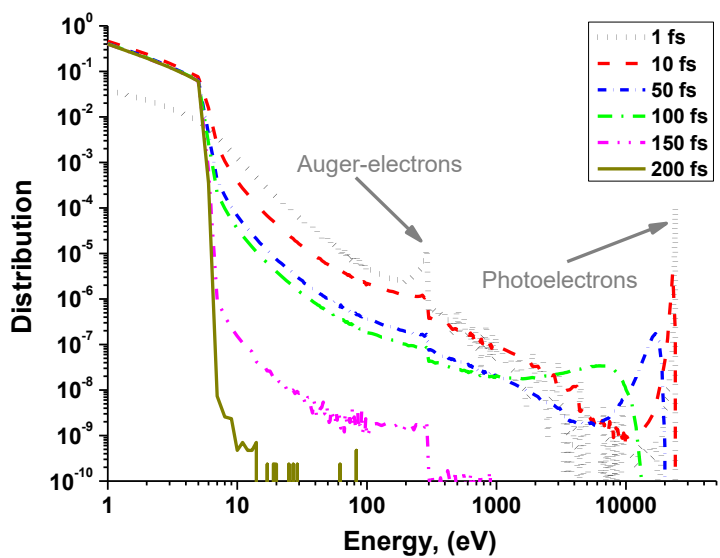

**Figure 2.** Evolution of the electron distribution function in diamond irradiated with an FEL pulse of 1 fs,

photon energy 24 keV, fluence of 1 mJ/cm$^2$.

Fig. 2 shows the transient electron distribution in diamond after irradiation with an FEL pulse (the

same parameters as for Fig. 1). The initial electron distribution is formed by electrons excited from the

valence and conduction bands or from deep shells of carbon atoms, shifted to high energies by the amount

of the photon energy (~24 keV) [45]. The secondary peak around 300 eV is formed by Auger-electrons

appearing due to decays of K-shell holes of carbon atoms which occur on a characteristic time of ~8 fs

[48]. Relaxation of the high-energy nonequilibrium tail of the distribution is predominantly governed by

the impact ionizations of secondary electrons during development of electron cascades. Duration of these

cascades depends on the initial excitation level (photon energy and fluence), and the material parameters

[46,47]. For the initial electron energy of ~24 keV in diamond, these cascades last ~120 femtoseconds

---

[†] N. Medvedev and B. Ziaja. XCASCADE, a Monte Carlo tool for modeling ultrafast electron cascading in solids after X-ray irradiation. DESY, Hamburg, 2014.

[46], as we also have seen in Figure 1.

Low-energy electrons are usually treated as non-interacting electron gas. More detailed studies, which included interactions between excited electrons, demonstrated partial thermalization of electrons with energies below ~10 eV occurring within a few femtoseconds [10,11,49]. This low-energy fraction of electrons follows then the Fermi-Dirac distribution with still increasing temperature being provided from interaction with high-energy nonequilibrium electrons [11,50].

Detailed analysis demonstrated that the transient nonequilibrium shape of the electron distribution in different materials (metals [11], semiconductors [45], insulators [47]) during and after an FEL pulse follows the so-called 'bump-on-hot-tail' distribution [51], similar to that shown in Figure 2. It consists of a large number of electrons populating low-energy states with (nearly) thermal distribution, while a minority of electrons remains in the long nonequilibrium tail of the distribution at high energies. Although the number of such high-energy electrons is small compared to those within the low-energy thermalized electron fraction, they possess rather large amount of energy, and therefore cannot be neglected and assumed to be thermalized at femtosecond timescales [11]. For highly excited warm-dense matter and plasma, this shape of the transient electron distribution was also confirmed by different methods: e.g. solutions of Boltzmann kinetic equation [10], and molecular dynamics simulations [49].

As it was shown in Ref. [11], different experimental techniques have access to different parts of the electron distribution function on femtosecond timescales. Thus, the 'bump-on-hot-tail' shape of transient nonequilibrium electron distribution must be taken into account for interpretation of experimental data [11,51]. For example, interpretations based on the assumption of thermal equilibrium produced contradicting results: a spectrum from radiative decay of L-shell holes in FEL-irradiated aluminum showed electron temperatures of ~1 eV [52], whereas Bremsstrahlung spectra from the same system were attributed to the electron temperatures of ~40 eV [53]. But such results could be explained by the fact that in the radiative decays L-shell holes are filled by electrons from the thermalized low-energy part of the transient electron distribution, while Bremsstrahlung spectra were formed by high-energy nonequilibrium tail of the 'bump-on-hot-tail' distribution [11].

## 2. SHI impact

The developed Monte Carlo code TREKIS[‡] [54] was used for modeling of electron kinetics in different materials after a swift-heavy ion impact (shown in Figure 1 above, and those discussed below). Transient electron distribution after an SHI impact demonstrates qualitatively similar behavior to the FEL-case [21,55]. Figure 3 presents the evolution of electron distribution in diamond after irradiation with a Ca ion with the energy of 11.4 keV/amu. Its initial shape is different from FEL case, because an ion excites electrons of different energies. Their spectrum scales approximately as $\sim 1/E^2$, except for low energies where it exhibits a plasmon peak [56].

This spectrum then quickly relaxes to the shape similar to the 'bump-on-hot-tail' distribution discussed above. However, since: (i) in the initial electron spectrum from SHI impact there is no peak similar to that of photoelectrons as after an FEL pulse, and (ii) a majority of primary electrons occupies low energy states, it makes the relaxation of the initial electron spectra somewhat faster in SHI-irradiation case (Fig. 3 vs Fig. 2). By the time of ~90 fs most of the electrons are already in the low-energy states (99% of them are below the ionization threshold), however, a minor amount of electrons up to ~30 eV is still present. Their number is so small that the secondary impact ionizations produced by them are almost not noticeable compared with the number of already excited electrons (the saturation regime in Fig. 1).

However, taking into account possible effects of the nonequilibrium electron distribution is important because it may affect interpretation of experiments: femtosecond diagnostics (< 100 fs) based on radiative photoemission or on Auger-electrons spectra [57], must keep in mind that they have access only to the low-energy part of the nonequilibrium distribution, and therefore estimated temperatures do not completely describe the excited electron ensemble. The shorter timescales a diagnostic tool probes,

---

[‡] N. Medvedev, R. Rymzhanov and A.E. Volkov, TREKIS: a Monte Carlo tool for modelling Time-Resolved Electron Kinetics in SHI Irradiated Solids. 2014.

the more pronounced are nonequilibrium effects.

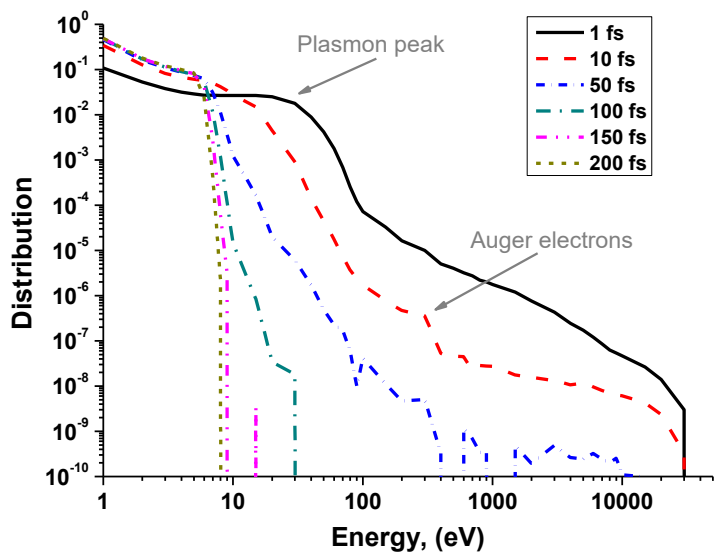

**Figure 3.** Evolution of the electron distribution function in diamond irradiated with Ca ion (457 MeV, or 11.4 MeV/amu).

Important difference between FEL and SHI irradiation occurs in the spatial distribution of the initial electronic excitations. In case of an SHI impact, homogeneous excitation cannot be assumed, i.e. periodic boundary conditions only along the ion-trajectory can be implemented in the cylindrical geometry [33,58]. Strongly localized initial excitation spreads out radially on femtosecond timescales, significantly changing radial profile of the electron density (and energy), as shown in Fig. 4. Initial ballistic transport of electrons gradually turns into diffusive propagation on the scales of a hundred femtosecond, alongside with the energy distribution relaxation [55,59]. But again, nonequilibrium part of fast electrons is important, since they possess a large amount of energy, and are spreading ballistically on femtosecond timescale, in contrast to diffusive motion of low-energy electrons [21,59].

The amount of energy kept by electrons with energies above a certain value is shown in Figure 5 as a function of time. We can see that at the beginning about 95 % of the energy provided by the SHI is accumulated in the electrons with energies above 10 eV, which are capable of performing secondary impact ionizations. Electrons with energies above 100 eV possess over 40 % of the energy forming the ballistic front shown in Fig. 4. Due to the secondary cascading, electrons lose their energy and the fraction of high-energy electrons quickly decreases. By the time of ~90 fs, less than 1% of the energy is kept by the electrons with energies above 10 eV. This indicates that most of the electrons are already slow, as we could also see in the electron spectrum, Fig. 3. Slow electrons exhibit diffusive behavior instead of the ballistic one [21,59], implying that in case of diamond irradiated with an SHI of 11.4 MeV/amu, after the time of ~90 fs a thermodynamic approach can be used for description of the excited electron ensemble.

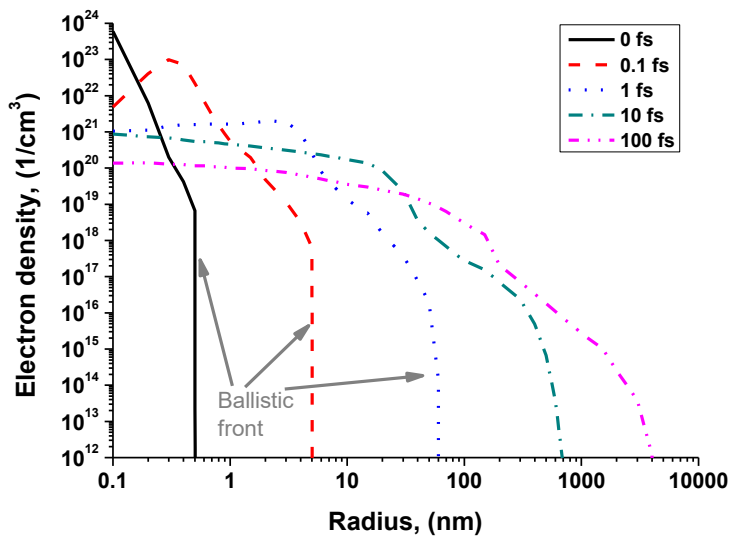

**Figure 4.** Radial electron distribution in diamond irradiated with Ca ion (457 MeV, or 11.4 MeV/amu).

It is also important to mention that many MC models are using a hard cut-off energy, stopping electrons as soon as their energy drops below a certain value (typically, around 10 eV). Out of this condition, the so-called total deposited dose vs radius is determined [32–34]. Obtained radial profile of the deposited dose is often used as an initial condition for further thermodynamic calculations, such as two-temperature model (see e.g. [60]). Such an approach, however, is inherently inconsistent, since neither the particle, nor the energy diffusion for low-energy electrons actually stops below any cut-off energy, therefore the spatial and temporal electron density evolution cannot be decoupled. This implies that the 'total dose' calculated in this manner does not correspond to any particular time instance, and, therefore, should not be used as the initial condition for further quantitative modeling [61]. An appropriate way to use the data characterizing the excited electron ensemble must: (a) account for the instance of electron thermalization, after which thermodynamic methods can be used, and (b) use transient electrons and energy distributions at a chosen time, without introducing artificial energy cut-offs [59,62].

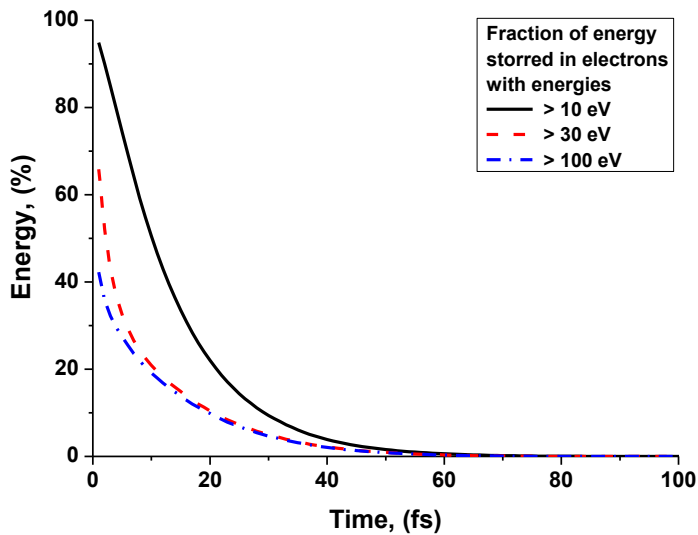

**Figure 5.** Fraction of energy possessed by the electrons with energies above a certain value in diamond irradiated with Ca ion (457 MeV, or 11.4 MeV/amu).

### III. Atomic dynamics stimulated by electron-lattice coupling

As we have seen in the previous section, within ~100 fs after FEL or SHI irradiation, highly excited electrons are relaxing into low-energy thermalized states. During and after this relaxation, these electrons are coupling to the lattice, which may stimulate structure transformations in a material. Electron-lattice coupling is dominated by two effects following from the transfer of kinetic and potential energies to atoms. The direct increase of the kinetic energy (temperature) of the lattice atoms is often referred to as the 'electron-phonon coupling' in the literature [63]. The change of the potential energy modifies the interatomic potential that may cause the so-called 'nonthermal melting' of lattice [64,65]. Below we will discuss these effects as well as the models and approximations with which they can be addressed. After that, we will give a few examples of atomic kinetics where the both effects and their interplay are investigated for conditions typical for SHI and FEL irradiation.

### 1. Thermal electron-lattice coupling

Before we present the models applicable to describe femtosecond electron-lattice energy exchange, let us start with the consideration of an often-used approximation for description of electron-lattice kinetic energy exchange: the phonon approximation for the atomic dynamics. In addition to splitting the atomic Hamiltonian into two parts describing the equilibrium atomic positions and their displacements (see discussion on Eq. (2) below), phonon approximation also assumes that [63]: (a) atomic motion can be well approximated by harmonic oscillations (which means that atoms are located close to the minimum of their potential); (b) a perfect periodicity in the atomic lattice allows to decouple such harmonic oscillators; (c) all atoms interact with their neighbors at least within their correlation length (phonon wave-length), so that their motion can be described in terms of collective variables [66]. These assumptions are usually satisfied for conditions realized for pico- and nano-second laser pulses. However, strictly speaking, after a femtosecond laser pulse irradiation, or an SHI impact, none of the approximations (a-c) is valid: the lattice can be heated to a high temperature or even nonthermally melted, thus, the interatomic potential is strongly unharmonic and quickly changing in time; they can be locally excited and 'melted' on a scale smaller than a phonon wave-length (especially in case of an SHI impact), which breaks the periodicity; interactions of atoms require times longer than the distance between them divided by the speed of sound – that makes a characteristic timescale of collective motion to be at least on the order of a few tens of femtoseconds (inverse phonon frequency). This is much longer than the femtosecond scales of processes considered for a femtosecond laser or SHI irradiation. Therefore, more general methods must be used to describe electron-lattice coupling on femtosecond time scales [31,66].

There are two standard approaches for describing interaction of the electrons with lattice in solids (and dense plasmas) beyond the phonon approximation: (i) free-electron approximation, coming from the plasma physics, and (ii) tight-binding approximation, borrowed from chemistry.

(i) The free-electron approximation assumes weak interaction between electrons and ions, when a gas of free electrons is scattering on a dynamically-coupled atomic ensemble. This allows to solve equations of motion for electrons and atoms independently accounting only for a minor corrections coming from their scattering events. This approximation results in Boltzmann kinetic equation for electrons where the electron-lattice interaction is included via electron-ion collision integral. The collision integral describing the kinetic energy transfer to the atomic subsystem depends on the *full* potential of electron-ion interaction, affected by the atom-atom dynamical and spatial correlations [67]:

$$I_{e-at}^{free} = -\frac{4}{(2\pi)^5 \hbar^2} \int d\mathbf{k}_i d\mathbf{k}_f \left|V(\mathbf{q})\right|^2 \left\{ f_{\mathbf{k}_f}(1 - f_{\mathbf{k}_i})S(-\mathbf{q}, -\omega) - f_{\mathbf{k}_i}(1 - f_{\mathbf{k}_f})S(\mathbf{q}, \omega) \right\}, \quad (1)$$

where $E_{\mathbf{k}} = \left(\hbar^2 \mathbf{k}^2\right)/2m_e$ is the kinetic energy of a free electron; $\boldsymbol{q} = \boldsymbol{k}_f - \boldsymbol{k}_i$ is the vector of momentum transferred by an electron during scattering from the initial ($\boldsymbol{k}_i$) to the final ($\boldsymbol{k}_f$) state; $V(\boldsymbol{q})$ is a Fourier-transform of the *full* electron-ion interaction potential; $f_k$ is the distribution function of the free-electron gas; and $S(\boldsymbol{q}, \omega)$ is the dynamic-structure factor (DSF) which is a Fourier-transform of the pair correlation function of the atomic ensemble [66].

The DSF formalism is actively employed for describing neutron scattering on solids [63,66,67], X-ray scattering [68–70], and electrons in liquid and amorphous materials [71–73] and metals [31,74]. For structured solids, it is often rewritten in terms of the complex-dielectric function [75] utilizing fluctuation-dissipation theorem [76].

(ii) In contrast, the tight-binding approximation assumes that electrons are closely following the atomic motion, which allows to split the atomic Hamiltonian into two parts [63]:

$$H_{at}(\{\mathbf{R}_i(t)\}) = H(\{\mathbf{R}_0\}) + \delta H(\{\delta \mathbf{R}_i(t)\}), \qquad (2)$$

where the equilibrium part of the atomic Hamiltonian $H(\{\mathbf{R}_0\})$ produces the equilibrium band-structure for electrons. After that, the electrons are assumed to be quasi-particles populating this band-structure (instead of a free-electron-gas dispersion curve). Interaction of such electrons with lattice realizes only via their interaction with atomic displacements $\delta\mathbf{R}_i$ from their equilibrium positions: $\delta H(\{\delta\mathbf{R}_i\})$.

Note that instead of the equilibrium atomic positions forming the lattice structure as assumed in the phonon approximation discussed above, one can trace evolution of the atomic positions step-by-step replacing $H(\{\mathbf{R}_0\})$ by $H(\{\mathbf{R}(t-\delta t)\})$.

It is worth to note that neglecting this so-called nonadiabatic interaction corresponds to the Born-Oppenheimer approximation for the electrons: electrons are then instantly readjusting on the new molecular orbitals that are evolving in time due to the atomic motion. This approximation is often used in *ab-initio* molecular dynamics codes such as DFT-MD. As was discussed in Ref. [30], Born-Oppenheimer approximation neglects electron-phonon (electron-ion) coupling, which can be described only beyond adiabatic effects.

Now, only the displacement part of the Hamiltonian enters the electron-lattice scattering within the tight-binding approximation instead of the *full* potential; the remaining part is 'compensated' by substituting electrons by quasi-particles on the new dispersion curves of the band-structure. Within this approximation, the electron-ion collision integral can be written as (see details in [30]):

$$I_{e-at}^{TB} = \frac{2\pi}{\hbar}\sum_{f=1}^{N}\left|M_{e-at}(E_i,E_f)\right|^2 \begin{cases} f_e(E_i)[2-f_e(E_f)]-f_e(E_f)[2-f_e(E_i)]g_{at}(E_i-E_f),\text{for } i > f \\ f_e(E_i)[2-f_e(E_f)]g_{at}(E_i-E_f)-f_e(E_f)[2-f_e(E_i)],\text{for } i < f \end{cases}, \qquad (3)$$

where electron distributions are now rewritten in the energy space instead of the momentum one (cf. Eq. (1)), populating the energy levels $E_i$ and $E_f$ (the eigenstates of the transient Hamiltonian $H(\{\mathbf{R}(t)\})$); $g_{at}(E)$ is the integrated atomic distribution (the integral of the Maxwellian distribution with a transient ion temperature), and $M_{e-at}(E_i,E_f)$ is the matrix element describing the scattering of an electron on the atomic displacement during the time-step $\delta t$, $\delta H(\{\delta\mathbf{R}_i(\delta t)\})$ [30]:

$$M_{e-at}(E_i,E_f) = \frac{1}{2}\left(\langle i(t-\delta t)|f(t)\rangle - \langle f(t-\delta t)|i(t)\rangle\right)\left(E_j-E_i\right). \qquad (4)$$

The matrix element depends on the overlap of the electronic wave functions (the eigenfunctions of the Hamiltonian) on the previous ($\left|i(t-\delta t)\right\rangle$ and $\left|f(t-\delta t)\right\rangle$) and the current time steps ($\left|i(t)\right\rangle$ and $\left|f(t)\right\rangle$).

This reflects the fact that the matrix element describes the electron transitions between the levels induced by the atomic displacements. In this formalism there is no distinction between the valence and the conduction band electrons, and all the energy levels (molecular orbitals) are addressed in the same unified manner.

## 2. Nonthermal melting

Nonthermal melting naturally occurs in any *ab-initio* Molecular Dynamics (MD) [64,65,77]. It is merely a result of the evolution of the atomic potential energy surface, which can be generally written as:

$$\Phi(\{\mathbf{R}(t)\},t) = \sum_{j=1}^{N} f_e(E_j)E_j(t) + E_{rep}(\{\mathbf{R}(t)\}),\tag{5}$$

where the attractive part is formed by the electrons via the electron distribution function $f_e(E_j)$ populating the transient energy levels $E_j$ (electron band structure) summed up over all these energy levels $N$; $E_{rep}$ is the core-core repulsive potential of ions.

Since the potential energy surface depends on the transient distribution of electrons, electronic excitation directly affects the interatomic potential [64,65,77]. When sufficient number of electrons is excited to antibonding states, the atomic potential is no longer attractive, and atoms start to experience new forces bringing them out of their former equilibrium positions. Note that in contrast to the Coulomb explosion, during the nonthermal melting the charge neutrality is preserved, and only the distribution function of electrons changes.

### 3. Examples: thermal and nonthermal melting under FEL irradiation

Below we will consider three examples of material damage under an FEL irradiation: (a) thermal melting of silicon [30]; (b) an interplay of thermal and nonthermal melting in silicon [30]; (c) purely nonthermal solid-to-solid phase transition: ultrafast graphitization of diamond [50,78,79].

As discussed in the previous section, the typical shape of the transient electron distribution function, the 'bump-on-hot-tail', allows simplifying the theoretical description of the nonequilibrium electron ensemble on femtosecond timescales [78]. The majority of low-energy electrons that are in near-thermal-equilibrium already at femtosecond time-scale can be described with the Fermi distribution. The minority of high-energy electrons far from equilibrium must be addressed with a nonequilibrium technique; in our case, we use individual particle Monte Carlo method of event-by-event simulations discussed above [30,78]. For tracing the atomic dynamics, we employ Molecular Dynamics (MD) on the time-dependent potential energy surface obtained from the transferable tight-binding Hamiltonian (see Ref. [78] for all the details), which naturally accounts for nonthermal effects. Nonadiabatic coupling between electrons and ions is additionally introduced within the tight-binding approximation according to Eqs. (3,4). This hybrid scheme is incorporated within the recently developed XTANT code [30].

Figure 6 presents the atomic structure of solid silicon irradiated with an FEL pulse under a normal incidence characterized by 1 keV photon energy, 10 fs duration (full width at half-maximum of the Gaussian temporal profile), and the photon fluence providing the absorbed dose of 0.7 eV/atom. As it was shown in Ref. [30], in this case the atomic dynamics is governed by the transfer of kinetic energy (heating of the lattice described by Eqs.(3,4)). This dose is just above the thermal melting of silicon ($E_{th}$ = 0.65 eV/atom). Fig 6. demonstrates that the phase transition to the transient low-density liquid state of silicon [80]. This state is characterized by the melted atomic structure, which has no long order in the atomic positions but with the kept local order (no amorphization) and semi-metallic electronic state (see [30]). Such melting occurs on a scale of a few ps, started by the fast electronic band gap collapse, while it takes a few picoseconds until electron and lattice temperatures fully equilibrate via heating of the atomic system due to nonadiabatic electron-ion coupling [30].

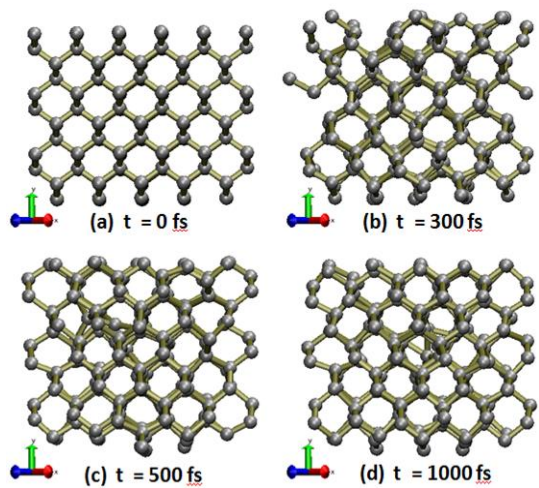

**Figure 6.** Atomic snapshots of the thermal melting of silicon under FEL irradiation of 1 keV photon energy, 10 fs pulse, 0.7 eV/atom absorbed dose. The figure is reproduced from Ref. [30].

In contrast, Figure 7 demonstrates nonthermal melting of silicon, irradiated with an FEL pulse with the higher photon fluence providing the dose of 1 eV/atom and the same other parameters as in Fig. 6. In this case, the absorbed dose is above the damage threshold leading to amorphization into high-density liquid [80] ($E_{th}$ = 0.9 eV/atom [30]). This amorphization occurs as an interplay between the thermal heating via electron-ion coupling (Eqs.(3,4)) and nonthermal melting resulted from modification of the interatomic potential (Eq.(5)). The interatomic potential is softened by the excitation of electrons when more than 4% of the valence band electrons are promoted to the antibonding states of the conduction band [30]. This results in the so-called 'phonon squeezing', where atoms are trying to adjust to the new potential energy surface formed by excited electrons [81]. Simultaneously, electron-ion coupling provides atoms with sufficient amount of the kinetic energy to overcome the reduced barrier for amorphization [30]. This nonthermal melting is very fast, and amorphization occurs already within 300-500 fs after the exposure to the laser pulse.

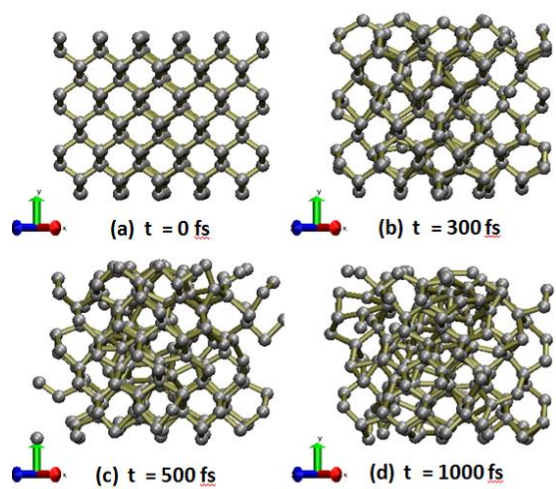

**Figure 7.** Atomic snapshots of the nonthermal melting of silicon under FEL irradiation of 1 keV photon energy, 10 fs pulse, 1 eV/atom absorbed dose. The figure is reproduced from Ref. [30].

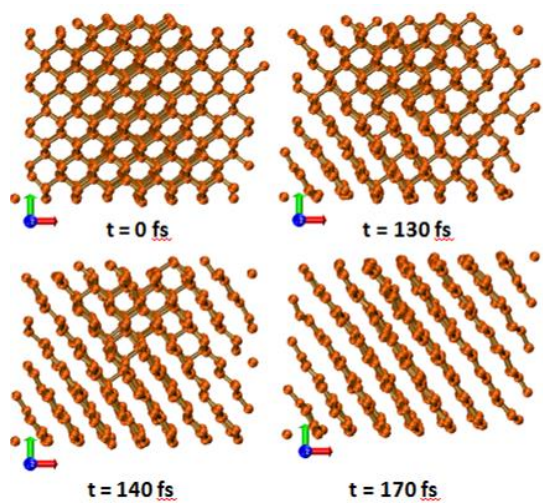

**Figure 8.** Atomic snapshots of the nonthermal graphitization of diamond under FEL irradiation of 10 keV photon energy, 10 fs pulse, 0.85 eV/atom absorbed dose. The figure is reproduced from Ref. [82].

Figure 8 demonstrates an example of a purely nonthermal phase transition: solid-to-solid phase transition from diamond to graphite. Diamond irradiated with FEL providing the dose of 0.85 eV/atom (graphitization threshold is $E_{th}$ = 0.7 eV/atom [78]) turns into graphite within some 50 fs after the energy is delivered to the low-energy (valence and conductions band) electrons. In the present example of 10 keV photon irradiation, the electron cascade takes around ~100 fs to bring the energy from high-energy part of the nonequilibrium distribution to the low-energy states of the valence and conduction band, as discussed in the section II. After that, the atoms experience a modified potential energy surface, and relax into the new phase of graphite. This happens extremely fast, much faster than nonthermal melting in silicon, because here the lattice is not required to amorphize, but only atoms to shift to the nearest positions corresponding to the new equilibrium solid state. As can be seen in Fig. 8, these are the positions in the graphite planes, while the $sp^3$-hybridization bonds are breaking. In experiments, diamond graphitization after exposure to a femtosecond FEL pulse was studied in Ref. [79], and more experiments are currently ongoing.

## 4. Examples: atomic dynamics after an SHI impact

In contrast to the X-ray FEL-case, when a solid is irradiated with a swift-heavy ion, strong gradients of excitation in the electronic system occur, affecting considerably the relaxation kinetics. To account for them, we used MC code TREKIS described above, combined with other approaches to model a track formation. An in-house hybrid code[§] couples TREKIS with another low-energy-electron kinetics tool (an approximate Boltzmann equation), and classical MD simulating the dynamics of excited lattice [83]. The DSF-formalism (Eq.(1)) is used to describe the electron-lattice coupling.

An example of the track kinetics in $Al_2O_3$ after Xe ion impact (167 MeV) was presented in [62]. It was shown that when electron coupling to the lattice was assumed within free-electron approximation for the conduction-band electrons appearing in a track (Eq.(1)), without contribution of valence-band holes, the resulting lattice heating seemed to be too low to produce experimentally observed damage regions, see Ref. [62]. Only taking into account the contribution from the excess energy of valence-band holes (in admittedly crude way), the track radius could be reproduced close to the experimental size [62]. The atomic snapshots are shown in Fig. 9 from the top-view. The importance of holes for the track kinetics seems to be coming from the fact that extremely fast electron transport out of the SHI trajectory carries energy away (as we saw in Fig. 3, and was discussed for a long time in the literature, see e.g. Refs. [3,31,84–86]). In contrast, valence holes are much slower (their kinetic energy is limited by the width of the valence band). Thus, these holes have enough time to interact with the atoms in the nanometric region around the ion trajectory providing them with energy sufficient to form a track.

---

[§] Developed by S. Gorbunov, P. Terekhin, N. Medvedev, R. Rymzhanov, and A. Volkov, 2013.

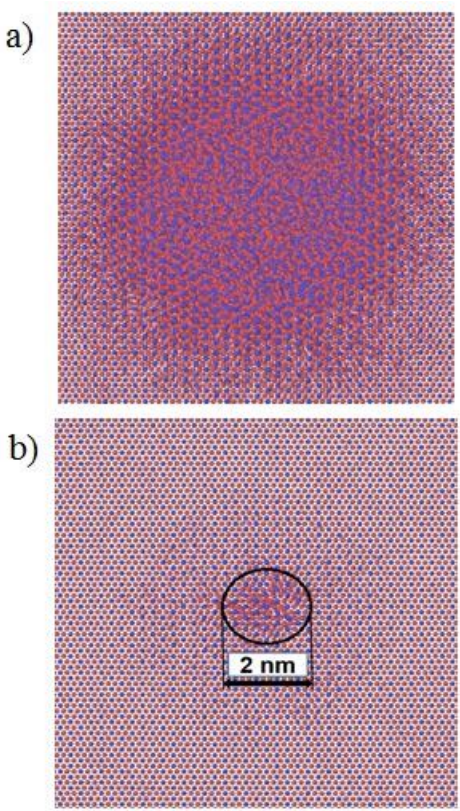

**Figure 9.** Atomic snapshots of $Al_2O_3$ after Xe ion impact (167 MeV). (a) 1 ps after the ion passage; (b) 50 ps after the ion passage. Al atoms are depicted as blue, O atoms are shown in red. Figure is reproduced from Ref. [62].

Contribution of valence holes to the lattice heating is often neglected in theoretical studies of SHI track formation. This results into a need for adjusting the input of the energy to the lattice by artificially changing parameters of the electron-phonon coupling to heat the lattice more significantly in the center before the electrons run away (see applications of the 'thermal-spike' model [60]). It might also be one of the reasons why rescaling of input data from TTM for MD simulations is often used [87,88], additionally to the specifics of the interatomic potentials, and the fact that softening of the potentials due to nonthermal effects is usually missing in classical-MD simulations. All of these effects and their relative importance warrant further dedicated research. Taking this into account, energy exchange between the lattice and valence holes seems to be necessary to avoid such fitting procedures. Another effect of valence holes that is often neglected is the Coulomb attraction of the excited electrons that might slow down their spreading. More detailed studies of these effects are currently in progress.

To the best of our knowledge, nonthermal melting in SHI tracks has not been studied quantitatively so far, although possibility of nonthermal melting is discussed in the literature [89–91]. Extreme levels of the initial electronic excitation, and their nanometric spatial localization resulting in strong spatial gradients inducing fast transport of particles and energy, pose additional challenges for theories, precluding direct application of previously developed schemes used for laser excitations. Moreover, when the spatial scales become smaller than the scales of the collective behavior of atoms (phonon wavelength), the two representations: real-space (coordinate), and reciprocal-space (momentum), are non-equivalent. That presents additional challenges for theoretical treatments of coupled electron and atomic dynamics in SHI tracks to be solved in future.

## IV. Summary

In this brief review we discussed similarities and differences in transient material excitation and evolution caused by free-electron-laser and swift-heavy-ion irradiation. The starting points for these kinds of irradiation are obviously different: photoabsorption delivers quanta of energy, whereas an ion impact produces many electrons with different energies (initial spectrum approximately follows $\sim 1/E^2$). Also, spatial profile of an FEL beam creates different geometry compared with an ion impact. In spite of these differences in the initial spectra of the energy deposition, the basic mechanisms of the subsequent electron relaxation processes are essentially the same. Coupling to the lattice and further atomic dynamics also have the same underlying physics. The relative importance of different channels of energy dissipation is determined by the geometry and initial spectrum of excitation in FEL vs SHI irradiation.

Taking this into account, two methods of description of electron-lattice coupling appropriate for FEL and SHI irradiation are discussed (free-electron approximation and tight-binding approach). Based on these methods, possibilities of thermal and nonthermal melting in FEL spots were demonstrated; an insufficient heating of the lattice solely by excited free electrons in SHI tracks is discussed. For spatially inhomogeneous excitations, strong gradients of density and energy-density pose challenges for theoretical descriptions that must be addressed in future modeling tools.

## Acknowledgements

The authors thank S. Gorbunov, H. Jeschke, Z. Li, O. Osmani, B. Rethfeld, R. Rymzhanov, R. Santra, K. Schwartz, P. Terekhin, C. Trautmann, O. Vendrell for their contribution to the previous works which were discussed throughout this review, and for numerous valuable discussions. Partial financial support from grants 13-02-1020, 15-02-02875, 15-58-15002 of Russian Foundation for Basic Research is acknowledged by A. E. Volkov.

## References

[1]     A. Ng, Int. J. Quantum Chem. 112 (2012) 150–160.
[2]     M.W.C. Dharma-wardana, F. Perrot, Phys. Rev. E. 58 (1998) 3705–3718.
[3]     A.E. Volkov, V.A. Borodin, Nucl. Instruments Methods Phys. Res. Sect. B Beam Interact. with Mater. Atoms. 107 (1996) 172–174.
[4]     A. Kaiser, B. Rethfeld, M. Vicanek, G. Simon, Phys. Rev. B. 61 (2000) 11437–11450.
[5]     T.G. White, J. Vorberger, C.R.D. Brown, B.J.B. Crowley, P. Davis, S.H. Glenzer, et al., Sci. Rep. 2 (2012) 889.
[6]     P. Lorazo, L. Lewis, M. Meunier, Phys. Rev. B. 73 (2006) 134108.
[7]     D.R. Rittman, C.L. Tracy, A.B. Cusick, M.J. Abere, B. Torralva, R.C. Ewing, et al., Appl. Phys. Lett. 106 (2015) 171914.
[8]     B. Rethfeld, A. Kaiser, M. Vicanek, G. Simon, Phys. Rev. B. 65 (2002) 214303.
[9]     V.P. Lipp, A.E. Volkov, M.V. Sorokin, B. Rethfeld, Nucl. Instruments Methods Phys. Res. Sect. B Beam Interact. with Mater. Atoms. 269 (2011) 865–868.

[10]  R.R. Fäustlin, T. Bornath, T. Döppner, S. Düsterer, E. Förster, C. Fortmann, et al., Phys. Rev. Lett. 104 (2010) 125002.

[11]  N. Medvedev, U. Zastrau, E. Förster, D.O. Gericke, B. Rethfeld, Phys. Rev. Lett. 107 (2011) 165003.

[12]  B. Rethfeld, A. Rämer, N. Brouwer, N. Medvedev, O. Osmani, Nucl. Inst. Methods Phys. Res. B. 327 (2014) 78–88.

[13]  W. Ackermann, G. Asova, V. Ayvazyan, A. Azima, N. Baboi, J. Bähr, et al., Nat. Photonics. 1 (2007) 336–342.

[14]  P. Emma, R. Akre, J. Arthur, R. Bionta, C. Bostedt, J. Bozek, et al., Nat. Photonics. 4 (2010) 641–647.

[15]  D. Pile, Nat. Photonics. 5 (2011) 456–457.

[16]  E. Allaria, R. Appio, L. Badano, W.A. Barletta, S. Bassanese, S.G. Biedron, et al., Nat. Photonics. 6 (2012) 699–704.

[17]  M. Nakatsutsumi, K. Appel, G. Priebe, I. Thorpe, A. Pelka, B. Muller, et al., Technical Design Report: Scientific Instrument, High Energy Density Physics (HED), Scientific, European X-Ray Free-Electron Laser Facility GmbH, Hamburg, Germany, 2014.

[18]  B.L. Henke, E.M. Gullikson, J.C. Davis, At. Data Nucl. Data Tables. 54 (1993) 181–342.

[19]  E.D. Palik, Handbook of Optical Constants of Solids, Academic Press, 1985.

[20]  K. Schwartz, a. E. Volkov, K.-O. Voss, M.V. Sorokin, C. Trautmann, R. Neumann, Nucl. Instruments Methods Phys. Res. Sect. B Beam Interact. with Mater. Atoms. 245 (2006) 204–209.

[21]  N.A. Medvedev, A.E. Volkov, N.S. Shcheblanov, B. Rethfeld, Phys. Rev. B. 82 (2010) 125425.

[22]  J.F. Littmark, J.P. Ziegler, U. Biersack, (1985).

[23]  D. Fink, ed., Fundamentals of Ion-Irradiated Polymers, Springer Berlin Heidelberg, Berlin, 2004.

[24]  F. Aumayr, S. Facsko, A.S. El-Said, C. Trautmann, M. Schleberger, J. Phys. Condens. Matter. 23 (2011) 393001.

[25]  K. Schwartz, A. Volkov, M. Sorokin, C. Trautmann, K.-O. Voss, R. Neumann, et al., Phys. Rev. B. 78 (2008) 024120.

[26]  M. Harmand, R. Coffee, M. Bionta, M. Chollet, D. French, D.M. Zhu, et al., Nat Phot. 7 (2013) 215–218.

[27]  R. Sobierajski, M. Jurek, J. Chalupský, J. Krzywinski, T. Burian, S.D. Farahani, et al., J. Instrum. 8 (2013) P02010–P02010.

[28]  S.-K. Son, L. Young, R. Santra, Phys. Rev. A. 83 (2011) 033402.

[29]  A.A. Baranov, N.A. Medvedev, A.E. Volkov, N.S. Shcheblanov, Nucl. Instruments Methods Phys. Res. Sect. B Beam Interact. with Mater. Atoms. 286 (2012) 51–55.

[30]  N. Medvedev, Z. Li, B. Ziaja, Phys. Rev. B. 91 (2015) 054113.

[31]  A.E. Volkov, V.A. Borodin, Nucl. Instruments Methods Phys. Res. Sect. B Beam Interact. with Mater. Atoms. 146 (1998) 137–141.

[32]  B. Gervais, S. Bouffard, Nucl. Instruments Methods Phys. Res. Sect. B Beam Interact. with Mater. Atoms. 88 (1994) 355–364.

[33]  E. Kobetich, R. Katz, Phys. Rev. 170 (1968) 391–396.

[34]  M.P.R. Waligórski, R.N. Hamm, R. Katz, O. Ridge, Int. J. Radiat. Appl. Instrumentation. Part D. Nucl. Tracks Radiat. Meas. 11 (1986) 309–319.

[35]  D.D. Ryutov, Multipulse effects in the damage to the LCLS reflective optics, in: S.G. Biedron, W. Eberhardt, T. Ishikawa, R.O. Tatchyn (Eds.), Opt. Sci. Technol. SPIE 49th Annu. Meet., International Society for Optics and Photonics, 2004: pp. 58–65.

[36]  N. Itoh, D.M. Duffy, S. Khakshouri, A.M. Stoneham, J. Phys. Condens. Matter. 21 (2009) 474205.

[37]  M.R. Islam, U. Saalmann, J.M. Rost, Phys. Rev. A. 73 (2006) 041201.

[38]  B. Ziaja, H. Wabnitz, F. Wang, E. Weckert, T. Möller, Phys. Rev. Lett. 102 (2009) 205002.

[39]  P.B. Allen, Phys. Rev. Lett. 59 (1987) 1460–1463.

[40]  L.D. Pietanza, G. Colonna, S. Longo, M. Capitelli, Eur. Phys. J. D. 45 (2007) 369–389.

[41]   N.S. Shcheblanov, T.E. Itina, Appl. Phys. A. 110 (2012) 579–583.

[42]   B.Y. Mueller, B. Rethfeld, Phys. Rev. B. 87 (2013) 035139.

[43]   C. Jacoboni, L. Reggiani, Rev. Mod. Phys. 55 (1983) 645–705.

[44]   T. Boutboul, a. Akkerman, a. Breskin, R. Chechik, J. Appl. Phys. 79 (1996) 6714.

[45]   N. Medvedev, B. Rethfeld, New J. Phys. 12 (2010) 073037.

[46]   N. Medvedev, Appl. Phys. B. 118 (2015) 417–429.

[47]   N. Medvedev, B. Ziaja, M. Cammarata, M. Harmand, S. Toleikis, Contrib. to Plasma Phys. 53 (2013) 347–354.

[48]   O. Keski-Rahkonen, M.O. Krause, At. Data Nucl. Data Tables. 14 (1974) 139–146.

[49]   S.P. Hau-Riege, Phys. Rev. E. 87 (2013) 053102.

[50]   N. Medvedev, H.O. Jeschke, B. Ziaja, Phys. Rev. B. 88 (2013) 224304.

[51]   D.A. Chapman, D.O. Gericke, Phys. Rev. Lett. 107 (2011) 165004.

[52]   S.M. Vinko, U. Zastrau, S. Mazevet, J. Andreasson, S. Bajt, T. Burian, et al., Phys. Rev. Lett. 104 (2010) 225001.

[53]   U. Zastrau, C. Fortmann, R. Fäustlin, L. Cao, T. Döppner, S. Düsterer, et al., Phys. Rev. E. 78 (2008) 066406.

[54]   N.A. Medvedev, R.A. Rymzhanov, A.E. Volkov, J. Phys. D. 48 (2015) 355303.

[55]   R.A. Rymzhanov, N.A. Medvedev, A.E. Volkov, Phys. Status Solidi. 252 (2015) 159–164.

[56]   A. Akkerman, M. Murat, J. Barak, Nucl. Instruments Methods Phys. Res. Sect. B Beam Interact. with Mater. Atoms. 321 (2014) 1–7.

[57]   G. Schiwietz, K. Czerski, M. Roth, F. Staufenbiel, P.L. Grande, Nucl. Instruments Methods Phys. Res. Sect. B Beam Interact. with Mater. Atoms. 226 (2004) 683–704.

[58]   W. Eckstein, Computer Simulation of Ion-Solid Interactions, Springer Berlin Heidelberg, 2011.

[59]   O. Osmani, N. Medvedev, M. Schleberger, B. Rethfeld, Phys. Rev. B. 84 (2011) 214105.

[60]   M. Toulemonde, C. Dufour, E. Paumier, Phys. Rev. B. 46 (1992) 14362–14369.

[61]   M. Murat, A. Akkerman, J. Barak, 2007 9th Eur. Conf. Radiat. Its Eff. Components Syst. 55 (2008) 2113–2120.

[62]   P.N. Terekhin, R.A. Rymzhanov, S.A. Gorbunov, N.A. Medvedev, A.E. Volkov, Nucl. Instruments Methods Phys. Res. Sect. B Beam Interact. with Mater. Atoms. (2015).

[63]   N.W. Ashcroft, D.N. Mermin, Solid State Physics, 1st ed., Cengage Learning, Boston, 1976.

[64]   P. Stampfli, K. Bennemann, Phys. Rev. B. 42 (1990) 7163–7173.

[65]   P. Stampfli, K. Bennemann, Phys. Rev. B. 46 (1992) 10686–10692.

[66]   L. Van Hove, Phys. Rev. 95 (1954) 249–262.

[67]   G. Baym, Phys. Rev. 135 (1964) A1691–A1692.

[68]   J. Chihara, J. Phys. F Met. Phys. 17 (1987) 295–304.

[69]   J. Chihara, J. Phys. Condens. Matter. 12 (2000) 231–247.

[70]   R.W. Lee, S.J. Moon, H.-K. Chung, W. Rozmus, H.A. Baldis, G. Gregori, et al., J. Opt. Soc. Am. B. 20 (2003) 770.

[71]   M.P. Tosi, M. Parrinello, N.H. March, Nuovo Cim. B. 23 (1974) 135–171.

[72]   N.H. March, M.P. Tosi, Atomic Dynamics in Liquids, Courier Corporation, Chelmsford, 1991.

[73]   V.T. Shvets, A.P. Fedtchuk, Phys. Scr. 52 (1995) 722–725.

[74]   S.A. Gorbunov, N.A. Medvedev, P.N. Terekhin, A.E. Volkov, Nucl. Instruments Methods Phys. Res. Sect. B Beam Interact. with Mater. Atoms. (2014).

[75]   J.-C. Kuhr, H.-J. Fitting, J. Electron Spectros. Relat. Phenomena. 105 (1999) 257–273.

[76]   R. Kubo, Reports Prog. Phys. 29 (1966) 255–284.

[77]   H. Jeschke, M. Garcia, K. Bennemann, Phys. Rev. B. 60 (1999) R3701–R3704.

[78]   N. Medvedev, H.O. Jeschke, B. Ziaja, New J. Phys. 15 (2013) 015016.

[79]   J. Gaudin, N. Medvedev, J. Chalupský, T. Burian, S. Dastjani-Farahani, V. Hájková, et al., Phys. Rev. B. 88 (2013) 060101(R).

[80]  M. Beye, F. Sorgenfrei, W.F. Schlotter, W. Wurth, A. Föhlisch, Proc. Natl. Acad. Sci. U. S. A. 107 (2010) 16772–6.

[81]  E. Zijlstra, A. Kalitsov, T. Zier, M. Garcia, Phys. Rev. X. 3 (2013) 011005.

[82]  N. Medvedev, V. Tkachenko, B. Ziaja, Contrib. to Plasma Phys. 55 (2015) 12–34.

[83]  S.A. Gorbunov, N.A. Medvedev, R.A. Rymzhanov, P.N. Terekhin, A.E. Volkov, Nucl. Inst. Methods Phys. Res. B. 326 (2014) 163–168.

[84]  M.I. Kaganov, I.M. Lifshitz, L.V. Tanatarov, Sov. Phys. JETP. 4 (1957) 173.

[85]  I.A. Baranov, Y.V. Martynenko, S.O. Tsepelevich, Y.N. Yavlinskii, Physics-Uspekhi. 31 (1988) 1015–1034.

[86]  Y. V. Martynenko, Y.N. Yavlinskii, Sov. At. Energy. 62 (1987) 93–97.

[87]  M.C. Ridgway, T. Bierschenk, R. Giulian, B. Afra, M.D. Rodriguez, L.L. Araujo, et al., Phys. Rev. Lett. 110 (2013) 245502.

[88]  T. Bierschenk, R. Giulian, B. Afra, M.D. Rodriguez, D. Schauries, S. Mudie, et al., Phys. Rev. B. 88 (2013) 174111.

[89]  D.M. Duffy, S.L. Daraszewicz, J. Mulroue, Nucl. Instruments Methods Phys. Res. Sect. B Beam Interact. with Mater. Atoms. 277 (2012) 21–27.

[90]  V.V. Stegailov, Contrib. to Plasma Phys. 50 (2010) 31–34.

[91]  M. Murat, A. Akkerman, J. Barak, Nucl. Instruments Methods Phys. Res. Sect. B Beam Interact. with Mater. Atoms. 269 (2011) 2649–2656.